\documentclass[pdflatex,sn-mathphys-num]{sn-jnl}

\usepackage{multirow}%
\usepackage{amsmath,amssymb,amsfonts}%
\usepackage{amsthm}%
\usepackage{mathrsfs}%
\usepackage[title]{appendix}%
\usepackage{xcolor}%
\usepackage{textcomp}%
\usepackage{manyfoot}%
\usepackage{booktabs}%
\usepackage{algorithm}%
\usepackage{algorithmicx}%
\usepackage{algpseudocode}%
\usepackage{listings}%
\usepackage{graphicx}  %
\usepackage{float}  %
\usepackage{subfigure}  %
\usepackage[percent]{overpic}  %
\usepackage{tikz}  %
\usepackage{hyperref} %
\usepackage{lineno} %
\modulolinenumbers[5] %

\theoremstyle{thmstyleone}%
\theoremstyle{thmstyletwo}%

\theoremstyle{thmstylethree}%

\begin{document}

\title[Article Title]{LaPON: A Lagrange’s-mean-value-theorem-inspired operator network for solving PDEs and its application on NSE}


\author[1]{\fnm{Siwen} \sur{Zhang}}

\author*[1,2]{\fnm{Xizeng} \sur{Zhao}}\email{xizengzhao@zju.edu.cn}

\author[1]{\fnm{Zhengzhi} \sur{Deng}}

\author*[1]{\fnm{Zhaoyuan} \sur{Huang}}\email{zhaoyuanhuang@zju.edu.cn}

\author[1]{\fnm{Gang} \sur{Tao}}

\author[1]{\fnm{Nuo} \sur{Xu}}

\author*[3]{\fnm{Zhouteng} \sur{Ye}}\email{zhoutengye@buaa.edu.cn}

\affil[1]{\orgdiv{Ocean College}, \orgname{Zhejiang University}, \orgaddress{\city{Zhoushan}, \postcode{316021}, \state{Zhejiang}, \country{China}}}

\affil[2]{\orgdiv{Ocean Research Center of Zhoushan}, \orgname{Zhejiang University}, \orgaddress{\city{Zhoushan}, \postcode{316021}, \state{Zhejiang}, \country{China}}}

\affil[3]{\orgdiv{Hangzhou International Innovation Institute}, \orgname{Beihang University}, \orgaddress{\city{Hangzhou}, \postcode{311115}, \state{Zhejiang}, \country{China}}}


\abstract{
Accelerating the solution of nonlinear partial differential equations (PDEs) while maintaining accuracy at coarse spatiotemporal resolution remains a key challenge in scientific computing. Physics-informed machine learning (ML) methods such as Physics-Informed Neural Networks (PINNs) introduce prior knowledge through loss functions to ensure physical consistency, but their ``soft constraints'' are usually not strictly satisfied. Here, we propose LaPON, an operator network inspired by the Lagrange's mean value theorem, which embeds prior knowledge directly into the neural network architecture instead of the loss function, making the neural network naturally satisfy the given constraints. This is a hybrid framework that combines neural operators with traditional numerical methods, where neural operators are used to compensate for the effect of discretization errors on the analytical scale in under-resolution simulations. As evaluated on turbulence problem modeled by the Navier-Stokes equations (NSE), the multiple time step extrapolation accuracy and stability of LaPON exceed the direct numerical simulation baseline at 8× coarser grids and 8× larger time steps, while achieving a vorticity correlation of more than 0.98 with the ground truth. It is worth noting that the model can be well generalized to unseen flow states, such as turbulence with different forcing, without retraining. In addition, with the same training data, LaPON's comprehensive metrics on the out-of-distribution test set are at least approximately twice as good as two popular ML baseline methods. By combining numerical computing with machine learning, LaPON provides a scalable and reliable solution for high-fidelity fluid dynamics simulation, showing the potential for wide application in fields such as weather forecasting and engineering design.
}




\maketitle



\section*{Introduction}

Nonlinear partial differential equations (PDEs) play a vital role in engineering and the physical sciences. They are used to describe and simulate complex physical systems, with applications ranging from electromagnetic field and communication research to vehicle and engine design, weather and climate prediction, plasma physics and quantum mechanics \cite{richardson1922weather, anderson1995computational, tang2005advances, messiah2014quantum}. Despite the direct relationship between these equations and the fundamental laws of physics, large-scale direct numerical simulation (DNS) is impossible. This fundamental problem has plagued the scientific computing community for decades because accurate simulations must resolve the smallest possible spatial and temporal scales, which imposes an unacceptable computational cost. For example, turbulent fluid flows \cite{pope2001turbulent}, which are fundamental to oceanographic, meteorological and fluid dynamics simulations, have very small minimum vortex sizes \cite{kolmogorov1995turbulence}. Traditional computational fluid dynamics (CFD) methods, including pseudo-spectral, finite element, and finite volume methods, rely on smoothly varying field representations, necessitating grids that capture even the finest features to ensure convergence. Instead, the traditional approach is to use simplified versions of the Navier–Stokes equations \cite{moser2021statistical, meneveau2000scale}, such as Reynolds-averaged Navier–Stokes (RANS) \cite{boussinesq1877theorie, alfonsi2009reynolds} and large eddy simulation (LES) \cite{smagorinsky1963general, lesieur1996new}, which allow for the use of coarser grids by sacrifice some accuracy, but the high computational cost is still quite a intractable problem. 

In recent years, with the rapid development of artificial intelligence (AI) technology, the end-to-end prediction approach based on pure machine learning (ML) has been used to solve the above problems \cite{kim2019deep, anandkumar2020neural, bhattacharya2021model, wang2020towards, lusch2018deep, erichson2019physics}. Thanks to the powerful fitting ability and extremely efficient parallel computing capability of ML models, it performs quite well in terms of both computing speed and implementation difficulty, and even surpasses the existing traditional methods in terms of accuracy in some areas \cite{bi2022pangu, Jumper2021AlphaFold, Abramson2024AlphaFold3}. Since the Fourcastnet \cite{kurth2023fourcastnet} and the PANGU \cite{bi2022pangu}, the end-to-end approach is still one of the most popular research directions in the field of meteorological forecasting. It is worth noting that ML models are data-driven approaches rooted in statistics. This approach has undoubtedly achieved great success when the amount of data is sufficient. However, an objective problem is that there is a serious shortage of effective data for model learning in many fields. This environment, which is the opposite of big data, is often called small sample, and the small sample problem often leads to serious generalization problems of ML models. For example, when trained ML models are exposed to new working conditions, their performance will deteriorate rapidly. The reason is that ML models generally have strong fitting capabilities by nature, and their outputs often have a high degree of freedom. If the amount of training data is insufficient and there are no appropriate constraints, ML models often experience optimization challenges in complex solution spaces.

It is generally accepted that the data and knowledge collaborative-driven modeling is the key to solving the problem of small sample \cite{raissi2019pinn, karniadakis2021piml, meng2022piml}. In addition, it can provide a certain degree of interpretability and reliability under unknown conditions (i.e., not within the scope of the training data) for ML models. To this end, a flurry of recent attempts to use knowledge to constrain ML models has sprung up. One major family of approaches treats the deep learning (DL) training process as a solution process, and the PDEs to be solved is encoded into the loss function of DL \cite{raissi2019pinn}. These Physics-Informed Neural Network (PINN) approaches implement a kind of constraints on the model that utilize prior knowledge through the loss function. These approaches are a type of unsupervised learning. When a DL model is trained, the solution function of its equations is well fitted by the model. While these new approaches are uncomplicated and universal, and have less data dependence than pure ML approaches, they usually do not reduce the computational cost, and unlike conventional prediction models, the trained model in these approaches only fits one particular solution of the equation, so the model cannot be generalized. In order to regain the characteristics of ``training once, applying to many places'' of the ML model, another major thrust separates the definite condition of the PDEs and uses it as part of the DL model input, so as to explicitly and flexibly control the DL model to output the different particular solution accordingly \cite{lu2021deeponet, li2020fourier, lusch2018deep}. These approaches usually derive some solution operators of PDEs and design some general network architectures based on them, which are a kind of constraints on the DL model architecture that utilize prior knowledge. This initial typical operator learning approach also has the advantages of easy to use and universal, but introduces the problem of data dependence again.

The above two approaches, which have been popular in recent years, consider how to integrate knowledge into AI from the perspectives of loss function and network architecture. As we all know, DL technology mainly consists of four modules: data, model, loss function and optimization program. How to reasonably integrate knowledge into these four modules has naturally become a key issue in the intelligent solution of PDEs and even the AI empowerment of traditional science. The knowledge learned by the model through training data itself or carefully crafted data enhancement is usually called observational bias, such as symmetry information; the relevant knowledge introduced through customized ML model architecture generally belongs to inductive bias, such as some physical laws or invariants, which can be expressed in a certain mathematical form; the knowledge given to the model in the form of loss function and optimization program is often called learning bias, such as some physical information that can be expressed in the form of integral, differential or fractional equations, and some intuitively feasible training strategies designed heuristically based on fuzzy concepts \cite{karniadakis2021piml, meng2022piml}. In addition, the embedding of these knowledge is sometimes not in conflict. It is often ideal to use these different principles together to design a hybrid approach specifically for a specific problem \cite{Nelsen2021Random, Mao2021DeepM, sirignano2020dpm}. 

\begin{figure}[!t]
	\centering
    
    


    \begin{overpic}[width=0.45\textwidth, angle=0]{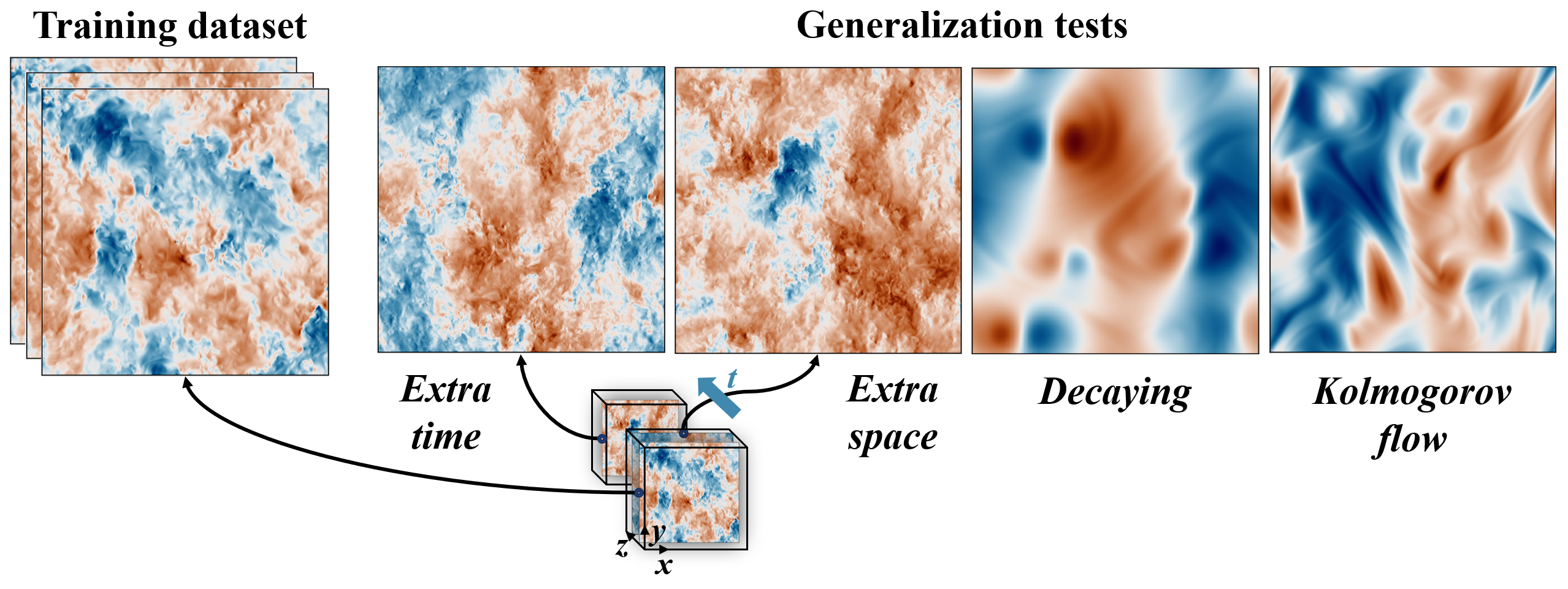} 
        \put(3,39){\textbf{(a)}}
    \end{overpic}
    \vspace{2mm}
    \begin{overpic}[width=0.45\textwidth, angle=0]{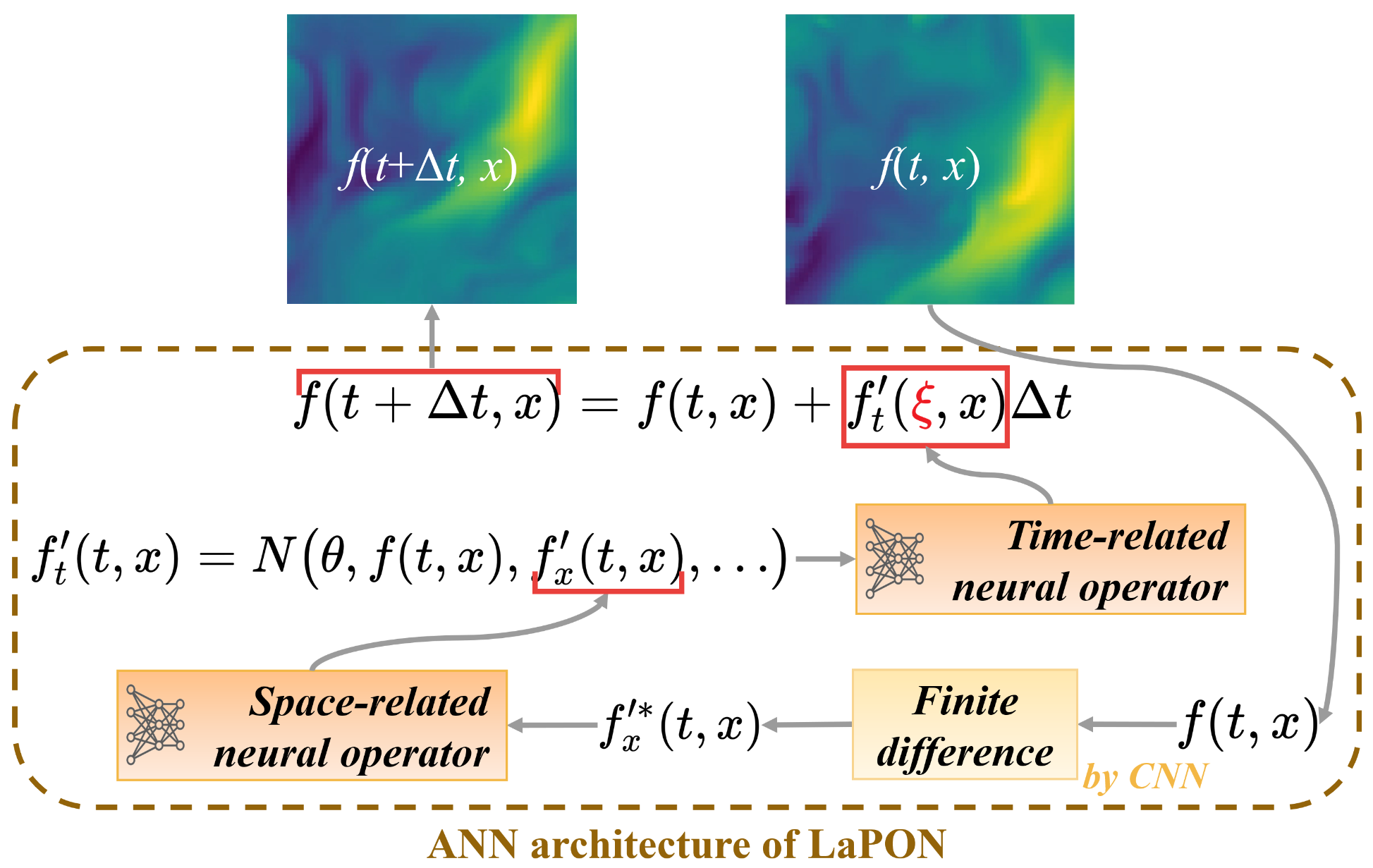}
        \put(3,62){\textbf{(b)}}
    \end{overpic}
    \vspace{2mm}
    \begin{overpic}[width=0.4\textwidth, angle=0]{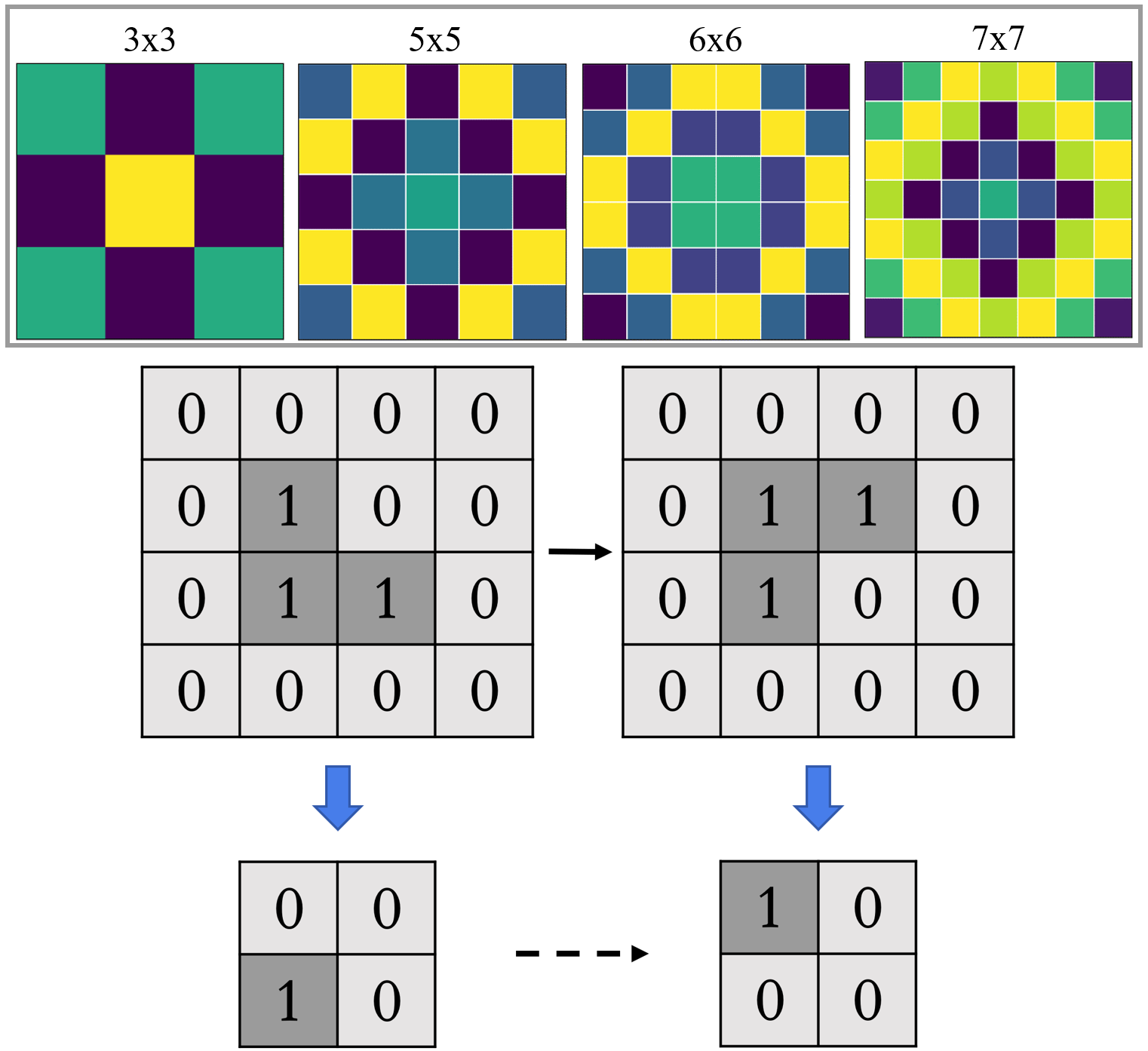}
        \put(-3,93){\textbf{(c)}}
    \end{overpic}
    
	\caption{Overview of our approach. (a) Illustrative examples of training and validation, showing the strong generalization capabilities of our model. (b) Structure of a single time step for our model, which integrates numerical computation and AI correction operators implemented by the convolutional neural networks. (c) Examples of isotropic convolution kernels of different sizes (in the solid box) and the corresponding rotational symmetry.} 
	\label{fig:main}
\end{figure}

Embedding knowledge in AI is essentially imposing constraints on it. Although the PINNs and the typical operator learning are uncomplicated and universal, both of them impose some kind of soft constraints on ML models. In other words, this form of constraint does not always work ideally, and the final constraint effect depends on factors such as the convergence of training in most cases. Therefore, sometimes we hope to design a corresponding hard constraints \cite{hendriks2020linearly, geist2020learning} on the neural network architecture for a specific problem, that is, no matter how the neural network parameters change, it will naturally strictly satisfy the constraints. Although this will face the problem of complex implementation, and its solution is usually not scalable and difficult to universal, it may still be the most principled approach at present. This type of approach can directly make the ML model strictly satisfy the given constraints, so that the model naturally has the advantages of reliable predictions on unseen scenarios and less data dependency. According to the characteristics of ML algorithms, it is relatively easy to implement hard constraints by customizing the model architecture. In current related research work, designing the DL model as a function or operator that satisfies a certain symmetry has become an important means to achieve hard constraints. This mainly includes embedding invariant \cite{ling2016reynolds, bloem2020probabilistic} in DL models and constructing equivariant neural networks \cite{hy2019covariant, tai2019equivariant, pfau2020ab, satorras2021n}. It is worth mentioning that such neural network architectures with built-in constraints are not new in the field of ML. Such design ideas have long been used in some classic models (such as the convolutional neural network, the long short term memory, the graph neural network, etc.). However, for AI applications in the field of computational physics, the development of constraints on neural network architecture and customized architecture design is still quite lacking. In addition, there are many advanced technologies that are relatively mature in the field of ML but are often overlooked when AI empowers traditional science, such as data enhancement technology that reduces model estimation error on limited samples, and various training strategies for reducing model optimization error. We will pay attention to the reference of some of these technologies in this work.

Here, we introduce a general method for computing the high-precision time evolution of solutions to the nonlinear time-dependent PDEs with high efficiency, using the coarser grid and the larger time step size than required by traditional numerical methods, and achieving the same accuracy. The method is based on a simple idea: to use AI to eliminate the discretization error of DNS under under-resolved simulations \cite{sirignano2020dpm, um2020solver, pathak2020using, kochkov2021machine} conditions, rather than giving the solution of the equation directly like end-to-end pure ML method. In addition, we also consider imposing additional constraints on the ML model, especially hard constraints. This work mainly verifies the effectiveness of the method based on the very challenging classical time-dependent PDE, Navier-Stokes equations (NSE), and related high-fidelity DNS turbulence database. Specifically, we propose a new solution method that corrects the partial derivative based on an operator in the process of numerical computation (rather than directly correcting the results of the numerical calculation) in a relatively more challenging time dimension, similar to the principle of the Runge-Kutta methods. And the idea provided by the Lagrange’s mean value theorem allows us to explicitly construct a time evolution computation form of the solution of PDEs in our neural network. Besides, in the spatial dimension, this approach does not average on unresolved sub-spatial scales like RANS; Instead, it utilizes operators to compensate for the effects of unresolved scales on resolved scales with low spatial resolution and low cost, similar to the concept of LES. The Lagrange’s mean value theorem (and the Taylor's theorem) show that the correction behavior of these operators is appropriate because the relevant residuals of traditional numerical methods are small, and their magnitude can be accurately estimated, which also makes subsequent constraints and ML model training possible. We use ML to discover these operators, called the neural operators, and impose the hard constraints based on the prior knowledge on them to ensure the reliability and low data dependence. The entire solution process of the algorithm is fully integrated into an ML model architecture, which enables it to have parallel computing capabilities and high computational efficiency consistent with the typical pure ML. The architecture is written as a differentiable program in a DL framework \cite{paszke2019pytorch} that supports reverse-mode automatic differentiation to optimize the above correction operators in the process of numerical computation. These AI operators learn how to calculate the fundamental features of variables related to the solution, rather than the typical end-to-end direct solution, and are therefore more easily generalized to different flow conditions, such as different flow intensities, different forcing, and even different Reynolds numbers and flow states. The model remains stable over long simulations (simulation over multiple time steps). Comparison with pure ML baselines verifies that the generalization mainly comes from the PDEs and physical constraints embedded in the ML architecture of the method.

\section*{Results}

\subsection*{PDE, datasets and data preprocessing}

\bmhead{Navier-Stokes}

Incompressible fluids are modeled by the dimensionless form of the Navier-Stokes equations: 
\begin{gather}
     \nabla \cdot \mathbf{u} = 0 \label{eq:con} \\
     \frac{\partial \mathbf{u}}{\partial t} = 
     -\nabla \cdot(\mathbf{u} \otimes \mathbf{u})
     +\frac{1}{Re} \nabla^{2} \mathbf{u}
     - \nabla p+\mathbf{f} \label{eq:ns} 
\end{gather}
where $t$ is the non-dimensional time, $\mathbf{u}$ is the non-dimensional velocity field, $\mathbf{f}$ is the non-dimensional external forcing, and $\otimes$ denotes a tensor product. The non-dimensional pressure $p$ can be obtained from the pressure Poisson equation (PPE) derived from Eq. (\ref{eq:con}). The Reynolds number $Re$ dictates the balance between the convection (first) or diffusion (second) terms in the right hand side of Eq. (\ref{eq:ns}). In convective-dominated flows, elevated Reynolds numbers amplify inertial effects over viscous dissipation, resulting in multiscale vortical structures that defy traditional modeling approaches. Such flows transition to turbulence when $Re \gg 1$ \cite{kochkov2021machine}, marking the onset of chaotic energy cascades.

\bmhead{Flow field datasets}

The main flow field dataset used for model training and validation in this work is the forced isotropic turbulence dataset in the Johns Hopkins Turbulence Database (JHTDB) \cite{li2008public}. The data is from a DNS of forced isotropic turbulence obeying the Navier-Stokes equation (Eq. (\ref{eq:ns})) at Taylor-scale Reynolds number Re = 418 on a periodic grid with 1024 spatial resolution, using a pseudo-spectral parallel code. After the simulation has reached a statistical stationary state, 5028 frames of data are generated. The duration of the stored data is about five large-eddy turnover times.

\bmhead{Data preprocessing}

It is an important issue in the DL field to appropriately scale the input data according to the characteristics of the DL model and to preprocess the training data reasonably according to the task type. In this work, in addition to the corresponding normalization operation of the input data (see Section \nameref{sec:nondimensionalization}), we also provide a strategy for dynamic data sampling to improve the training efficiency of the model and its performance at large discrete scales.

Specifically, during the training process, we adaptively adjust the spatiotemporal resolution of flow field sampling according to the iterative progress of model optimization. In the early stage of training, a smaller discrete scale is used as the main method, and as the training progresses, the time step size and grid size of the training data are gradually increased. At the same time, in order to increase the robustness of the model at different resolutions, we also randomly crop and scale the solution domain of the training data within a certain range to achieve multi-scale training of the model.

\subsection*{Model architectures}

In this work, we propose a method for computing high-precision time evolution of solutions to the nonlinear time-dependent PDEs in the under-resolved case. Compared with DNS, this method has a much looser requirement for the accuracy of spatial and temporal discretization. Based on the Lagrange’s mean value theorem, we design a neural operator to correct the standard DNS solution process for PDEs in the case of under-resolved simulations. Specifically, the method uses neural operators to perform residual correction on the numerical calculation process (rather than directly on the results) to compensate for the discretization error that objectively exists between the numerical solution and the exact analytical solution. This can be regarded as a form of coupled solution of AI and traditional numerical methods. Both the numerical calculation process and the AI correction operator are effectively integrated into one new neural network architecture in the form of a neural network module to ensure that the extremely high computational efficiency of the ML model is not lost (as shown in Fig. \ref{fig:main} (b)). The model is named LaPON (an Operator Network for PDEs solving based on the Lagrange’s mean value theorem). At the same time, the architecture maintains the differentiability of the entire network so that the learnable parameters of the neural operators in the architecture used to learn the fundamental features can be optimized using standard back-propagation techniques. These neural operators are also subject to the hard constraints based on the prior knowledge in the form of a customized design of the network architecture, where the constraints on the output of the neural operators will ensure that the lower limit of the model accuracy is no less than the first-order accuracy of the numerical method as much as possible.

\bmhead{Embedding PDEs into the network architecture}

Traditional numerical methods generally calculate the discrete numerical format of PDEs on the CPU. However, CPUs usually have a small number of cores, which makes their computational efficiency in some cases much lower than that of various accelerator devices including GPUs that have large-scale data parallel computing capabilities. With the popularization of high-performance computing resources, large-scale simulations using accelerator devices such as GPUs are becoming more and more important. In this work, in order to greatly improve the efficiency of numerical calculations and maintain the integration of the entire solution process to facilitate operator optimization, we directly embed the discrete numerical format of PDEs in the computation graph (directed acyclic graph) of a neural network. This is a relatively easy solution for simulation using accelerator devices. From another point of view, this operation realizes the direct embedding of knowledge into the neural network model architecture, which gives physical meaning to the computation of the neural network skeleton, improves its reliability and reduces the dependence of the trainable module on data.

Specifically, taking the NSE used in this work as an example, we use the explicit time marching method to preliminarily solve it. The equations are discretized using time forward differences and spatial central differences, and the numerical format is implemented using the DL framework, which constitutes the PDE module in the neural network architecture. Since the convolution operation of neural networks is closely related to finite difference in mathematical form (there is a corresponding relationship between the differential order of the convolution kernel/filter and the order of the differential operator) \cite{cai2012image, dong2017image}, all difference operations in this solution are implemented in the convolution layer by specifying weights. This allows the difference of the physical field to be efficiently calculated in parallel directly on the accelerator device.

\bmhead{Correcting discretization error}

\textbf{The discretization error} objectively exists in traditional numerical methods and is the main source of error when simulating on coarse grids. This is caused by the discretization of the solution domain and equations, and is inevitable in traditional numerical methods \cite{sirignano2020dpm, duraisamy2020machine}. In the context of the finite difference, discretization error refers to the truncation error of the Taylor formula. Its corresponding discretization scale is usually one of the important parameters that most significantly affects simulation accuracy and speed. In this work, a differential correction neural operator module is constructed and used to eliminate the residual between the numerical solution (here refers to the result obtained based on the finite difference of low order accuracy on a coarse grid and a large time step size) and the exact analytical solution (approximated by the result of the finite difference of high order accuracy on a fine grid and a small time step size in this work), which is the key to solving PDEs with high accuracy at a larger discrete scale. Among them, the mean value point idea provided by the Lagrange's mean value theorem allows us to explicitly construct the time evolution computation form of the solution of PDEs (the top layer of the neural network) in the neural network. In this way, the strict dependence of traditional numerical methods on tiny time steps can be broken.

Taking the nonlinear time-dependent PDE in two-dimensional space-time as an example, its general form is:
\begin{equation}
f_{t}^{\prime}(t, x)=N \left( \theta, f(t, x), f_{x}^{\prime}(t, x), f^{\prime \prime}_{ x x} (t, x), \ldots \right)
\end{equation}
where $N$ is a nonlinear operator and $\theta$ is a system parameter. Assuming that there is a solution function $f(t, x)$ on the solution domain $[a, b]$ for the PDE, which is continuous on the closed interval $[a, b]$ of the solution domain and differentiable on the open interval $(a, b)$. Then the Lagrange's mean value theorem shows that in the solution domain, for any spatial position $x_0$, any time $t_0$ and $t_0+\Delta t$, there exists at least one mean point $\xi \in (t_0, t_0+\Delta t)$, such that the function satisfies the following relationship:
\begin{equation}
f(t_0+\Delta t, x_0)=f(t_0, x_0)+f^{\prime}_t(\xi, x_0) \Delta t
\end{equation}

In the explicit time marching method of PDEs, the time derivative $f^\prime_t(\xi, x_0)$ at the mean point $\xi$ is approximated by $f^\prime_t(t_0, x_0)$ (in the implicit method, it is approximated by $f^\prime_t (t_0+\Delta t, x_0))$. This is one of the sources of the discretization error in the traditional numerical method in time marching. Here we define the corresponding residual of the time derivative as:
\begin{equation}
r_t(t_0, x_0):=f_{t}^{\prime}(\xi, x_0)-f_{t}^{\prime}(t_0, x_0)
\end{equation}

Although the Lagrange’s mean value theorem shows that the mean point must exist and there is at least one, there is currently no practical first-principles solution that can directly find the mean point and its derivatives in the process of solving PDEs. Fortunately, by training an ML model on enough data, accurately fitting the residuals can be achieved. Therefore, we define a neural operator $T$ related to the time derivative of the solution function and expect it to satisfy:
\begin{equation}
\left(T_{\mathbf{p}} f^{\prime}_t\right)(t_0, x_0)=r_t(t_0, x_0)
\end{equation}
where $T_{\mathbf{p}}$ represents the state of the operator $T$ working under the conditional parameter group $\mathbf p$ (the conditional parameters are used to control the behavior of the operator, so we call this type of operator the conditional neural operator), and the elements of the vector $\mathbf p$ include the time step $\Delta t$ and important characteristic parameters of the physical field (such as the Reynolds number Re of the flow field). Modeling the residual is appropriate both from the perspective of the Lagrange’s mean value theorem and the Taylor's theorem, and pragmatically because the relative residual between $f^\prime_t (\xi, x_0)$ and $f^\prime_t (t_0, x_0)$ in a single time step is small, and its magnitude can be accurately estimated, which also enables subsequent training and implementation of constraints. Therefore, the complete form of the neural network architecture based on the explicit time marching method and the Lagrange’s mean value theorem can be simply expressed as:
\begin{equation}
ANN(f(t_0, x_0)):=f(t_0, x_0)+\left(f^{\prime}_t(t_0, x_0)+\left(T_{\mathbf{p}} f^{\prime}_t\right)(t_0, x_0)\right) \Delta t
\end{equation}
so we have:
\begin{equation}
f(t_0+\Delta t, x_0)=ANN(f(t_0, x_0))
\end{equation}

If we consider only solving a time-dependent ordinary differential equation (ODE), the method is complete. For PDEs with spatial dimensions, we will face another problem: in the process of numerical solution, the exact $f^{\prime}_t (t_0, x_0)$ is approximated by the $f^{\prime *}_t (t_0, x_0)$ calculated by the spatial difference, which is another source of the discretization error. Therefore, we need another conditional neural operator $S$ related to the spatial derivative of the solution function of the PDE, which should satisfy:
\begin{equation}
\left(S_{\Delta x} f_{x}^{\prime *} \right)(t_0, x_0)=r_{s}(t_0, x_0):=f_{x}^{\prime}(t_0, x_0)-f_{x}^{\prime *}(t_0, x_0)
\end{equation}
where $ S_{\Delta x} $ represents the state of the operator $S$ when the spatial discrete scale is $\Delta x$, $r_s(t_0, x_0)$ is the residual of the spatial derivative, $ f^\prime_x(t_0, x_0) $ is the exact value of the spatial derivative of the solution function, and $f^{\prime*}_x(t_0, x_0)$ is the numerical approximation of the uncorrected spatial derivative of the solution function from the finite difference on a coarse grid. Therefore:
\begin{equation}
f_x^{\prime}(t_0, x_0)=f_x^{\prime *}(t_0, x_0)+\left(S_{\Delta x} f_{x}^{\prime *}\right)(t_0, x_0)
\end{equation}

Higher-order derivatives can be calculated iteratively based on this. Finally, $f^\prime_t (t_0, x_0)$ can be obtained by substituting these derivatives into the general form of PDE.

\subsection*{Imposing hard constraints based on prior knowledge}

\bmhead{Constraining the neural operator}

In short, the time-related operator corrects the derivative at the current moment to the derivative at the moment of the mean point described by the theorem according to the Lagrange’s mean value theorem. The space-related operator can be understood as being used to fit the remainder in the Taylor's theorem, correcting the spatial numerical derivative to an exact value. In order to make the operator effective, it is indispensable to train the ML model on a sufficient amount of data. Although the neural operator in this work has good universality as a basic module, so that a basic mapping can be learned on different physical field data, the lack of available training data in traditional scientific fields has been an objective and universal problem. As a data-driven model, we cannot expect it to produce ideal results on little data. Therefore, adding additional constraints to the neural operator to achieve the data and knowledge collaborative-driven modeling, thereby reducing the data dependence of the ML model, is an issue that needs to be considered. In addition, the reasonable imposition of additional constraints, especially hard constraints, can also make the ML model more reliable in unseen situations.

Specifically, we mentioned above that the residual that needs to be corrected by the neural operator has a provable and exact range, that is, there is an exact estimate of the error. Therefore, we can impose a hard constraint on the output of the neural operator to ensure that the final output is within this range, thereby ensuring the accuracy of the model as much as possible.

For time-dependent operators, we first assume that the convexity of the solution function of the PDEs is constant on the minimum time scale that the ML model can resolve. After determining the sign of the second-order time derivative at the current moment, we can know the magnitude relationship between $ f^{\prime}_t (t_0, x_0) $ and $ f^{\prime}_t (\xi, x_0) $, so the sign of $(T_p f^{\prime}_t)(t_0, x_0)$ will be determined. We can use the ReLU function as a constraint function to constrain the sign of the output value of the original operator module, so the module can be modified to a nested form of the original operator plus an additional constraint function:
\begin{equation}
r_t(t_0, x_0)=\operatorname{sign}\left(f^{\prime \prime}_{t t}(t_0, x_0)\right) * \operatorname{ReLU}\left(\operatorname{norm}\left(\left(T_{\mathbf{p}} f^{\prime}_t\right)(t_0, x_0)\right)\right)
\end{equation}
where $sign$ is the sign function, $norm$ is the normalization operation in ML, which is usually added before the activation function so that the activation function has better performance and the model has better convergence during training.

For the spatial related operator, assuming that the error of the nth order accuracy spatial numerical derivative $f^{\prime * (n)}_x(t, x)$ of the solution function $f(t, x)$ of the PDEs is much larger than the n+1th order (n is 1 in this work), that is, $|r_n| \gg |r_n+1|$, where $r_n$ represents the error of the nth-order precision numerical derivative, then there will obviously be a certain interval $[I_L, I_H]$ such that $f^{\prime}_x(t, x) \in [I_L, I_H]$ (see Section \nameref{app:proof} for the proof). Here, we take the mean point $I_C= f^{\prime *}_x (n+1)(t, x)$, and the radius $R=| f^{\prime * (n+1)}_x(t, x) - f^{\prime * (n)}_x(t, x) |$. Therefore, the sigmoid function can be used as a constraint function to constrain the range of the output value of the original operator module, so the module can also be modified to a nested form of the original operator plus an additional constraint function:
\begin{equation}
r_s(t_0, x_0)=\left(\operatorname{sigmoid}\left(\operatorname{norm}\left(\left(S_{\Delta x} f^{\prime * (n+l)}_x \right)(t_0, x_0)\right) / 2R\right)-0.5\right) * 2R
\end{equation}

We additionally constrain the corrected derivative to be within the interval $[I_L, I_H]$, which guarantees that the correction is at least nth order accurate, because its upper bound of error is only equal to the error of the space numerical derivative in the nth order accuracy. In addition, it is worth noting that we add an additional calculation of division by $2R$ to the output of the neural operator in the sigmoid function, which corresponds to the multiplication of $2R$ outside the function. This operation does not change the range of the original function, but it can scale the maximum value of the gradient of the learnable parameter back to the position before the constraint is imposed to avoid gradient diffusion and gradient explosion, thereby preventing the convergence of ML model optimization from being adversely affected by the additional constraints.

\bmhead{Constraining the basic operations}

The neural operators in this work are built on convolutional neural networks (CNNs), and convolution operations are their most basic modules. Usually, the constraints imposed on the basic operation will directly benefit the entire model. In fact, CNNs is an equivariant neural network. As an operation that captures fundamental information, convolution naturally satisfies translational symmetry, which enables CNNs to give consistent operation results for the translation transformation of data, which is exactly what we need. It is a basic prior assumption that physical quantities (such as flow fields) have spatial translational symmetry, which corresponds to the law of conservation of momentum. In order to obtain better model reliability and less data dependence, in this work, we also impose another additional constraint on the ML model: isotropic convolution.

Based on the spatial rotational symmetry (corresponding to the assumption of conservation of angular momentum), we designed a new convolution operation as the basic module of our ML model to ensure that the model gives consistent calculation results for the rotational coordinate transformation of the physical field. This convolution differs from the traditional convolution only in the convolution kernel: the learnable parameters of the convolution kernel are isotropic, that is, the learnable parameters are different only in their distance from the center of the convolution kernel, but are consistent in different directions (see Fig. \ref{fig:main} (c)). For such a convolution operation $W$, under any rotation transformation $R_\theta$, $W(R_\theta(x))=R_\theta(W(x))$ is satisfied. We achieve this by sharing weights and designing an algorithm to generate isotropic convolution kernels to serve the subsequent modeling of neural operators. By building neural operators with isotropic convolution, we can embed the prior knowledge of the conservation of angular momentum into our ML model.

\subsection*{Performance of LaPON}

\begin{figure*}[!ht]
	\centering 
	\includegraphics[width=1.0\textwidth, angle=0]{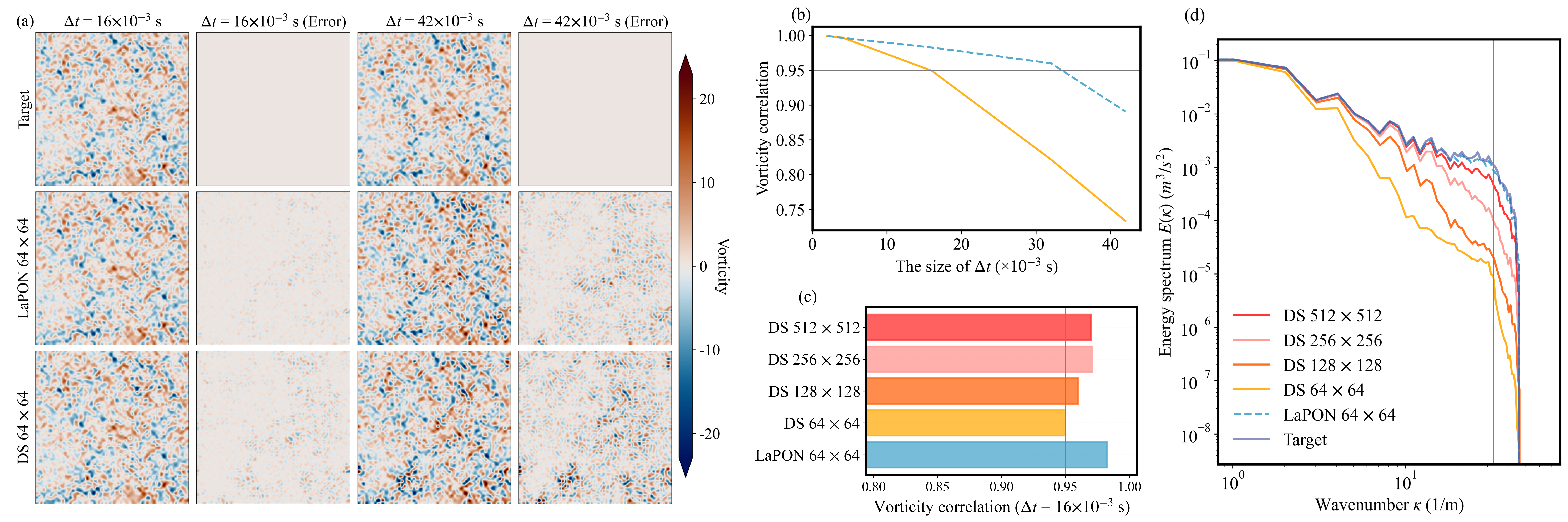}	
	\caption{LaPON exceeds accuracy of direct simulation at $8 \times $ higher spatial resolution, and the accuracy is also significantly higher on large time steps. (a) The vorticity field solved by our model (LaPON $64 \times 64$) and baseline (DS $64 \times 64$) on $8 \times $ and $21 \times $ base time step, and the corresponding error. (b) Comparison of the vorticity correlation between predicted flows and the reference solution for our model and direct simulation on different time step sizes. (c) Comparison of the vorticity correlation between predicted flows and the reference solution for our model and direct simulation on different spatial resolutions. (d) Energy spectrum on the first one Kolmogorov time (210 original time steps) using $1 \times $ the base time step. The gray vertical line indicates the effective wavenumber cutoff region of $64 \times 64$ resolution.} 
	\label{fig:acc}
\end{figure*}

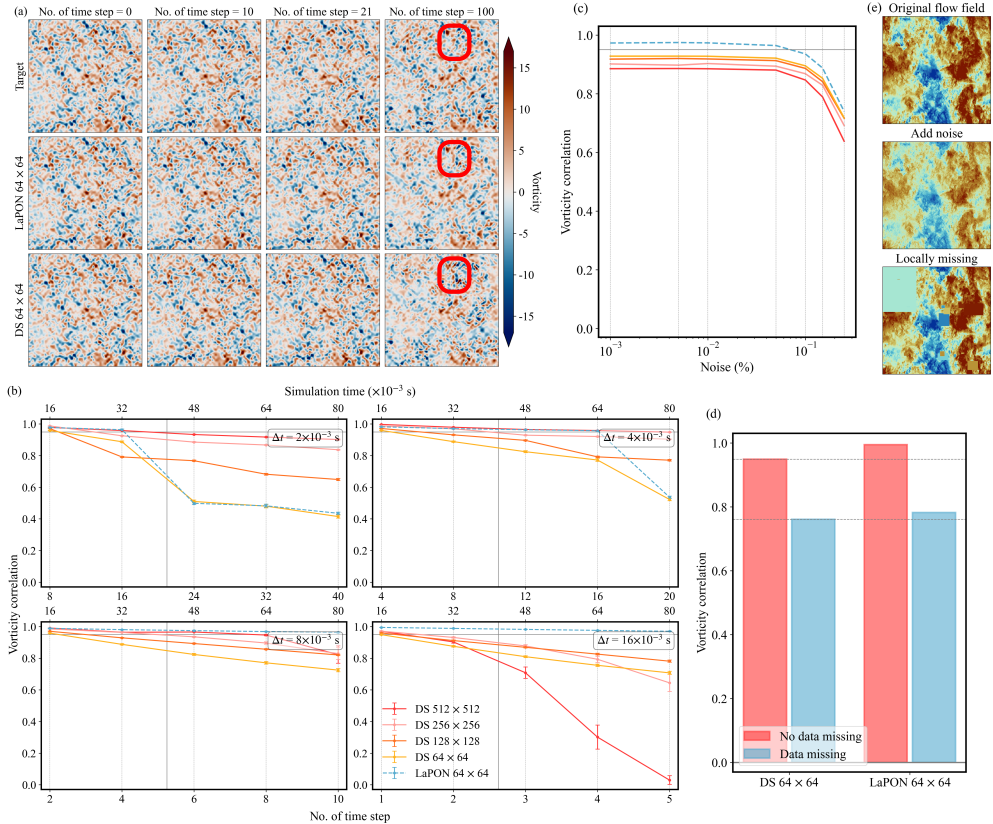
\begin{figure*}[!t]
	\centering 
    \begin{tikzpicture}
        \node[anchor=south west,inner sep=0] (image) at (0,0) {
        \includegraphics[width=1.0\textwidth, angle=0]{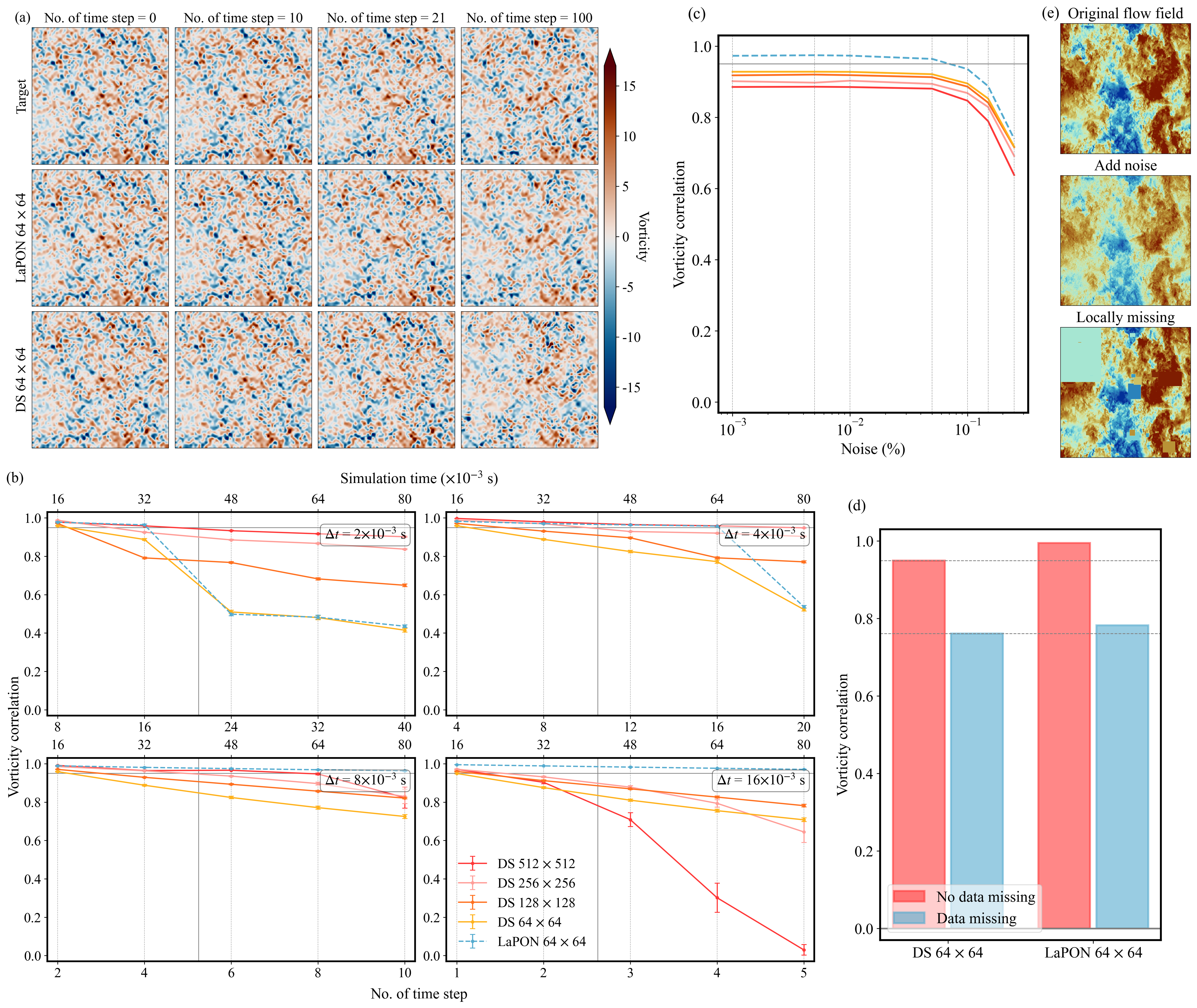}
        };
        \begin{scope}[x={(image.south east)},y={(image.north west)}]
            \draw[red,ultra thick,rounded corners] (0.44,0.925) rectangle (0.47,0.965);
            \draw[red,ultra thick,rounded corners] (0.44,0.785) rectangle (0.47,0.825);
            \draw[red,ultra thick,rounded corners] (0.44,0.645) rectangle (0.47,0.685);
        \end{scope}
    \end{tikzpicture}
	\caption{LaPON maintains accuracy and stability and exceeds direct simulation at $8 \times $ higher spatiotemporal resolution on large time steps that do not even satisfy the CFL condition. (a) Time evolution of the vorticity field  (using $1 \times $ the base time step) and the corresponding reference solution. (b) Comparison of the mean and standard deviation (in the sample dimension) of the vorticity correlation between the prediction of our model and the direct simulation baselines and the reference solution. The gray vertical solid line represents the first one Kolmogorov time. (c) Comparison of the noise immunity for our model and direct simulation baselines on different noise level (using $8 \times $ the base time step). (d) Compare the robustness of our model to the direct simulation baseline against partial data missing (using $8 \times $ the base time step). (e) Visualization of data noise and local missing.} 
	\label{fig:stb}
\end{figure*}

\begin{figure*}[!t]
	\centering 
	\includegraphics[width=1.0\textwidth, angle=0]{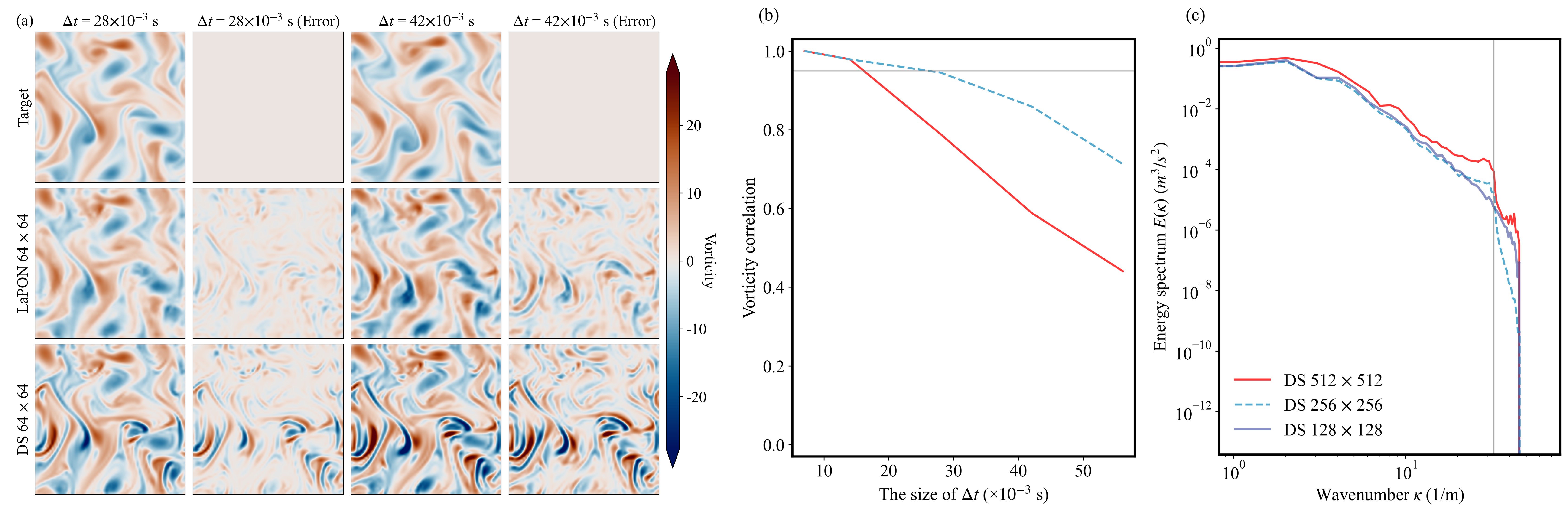}	
	\caption{LaPON generalizes well to decaying turbulence without retraining. (a) Prediction and error of vorticity fields. (b) Vorticity correlation between predicted flows and the reference solution. (c) Energy spectrum on the first 210 original time steps using $1 \times $ the base time step.} 
	\label{fig:gnr}
\end{figure*}

\begin{figure}[!t]
	\centering 
	\includegraphics[width=0.45\textwidth, angle=0]{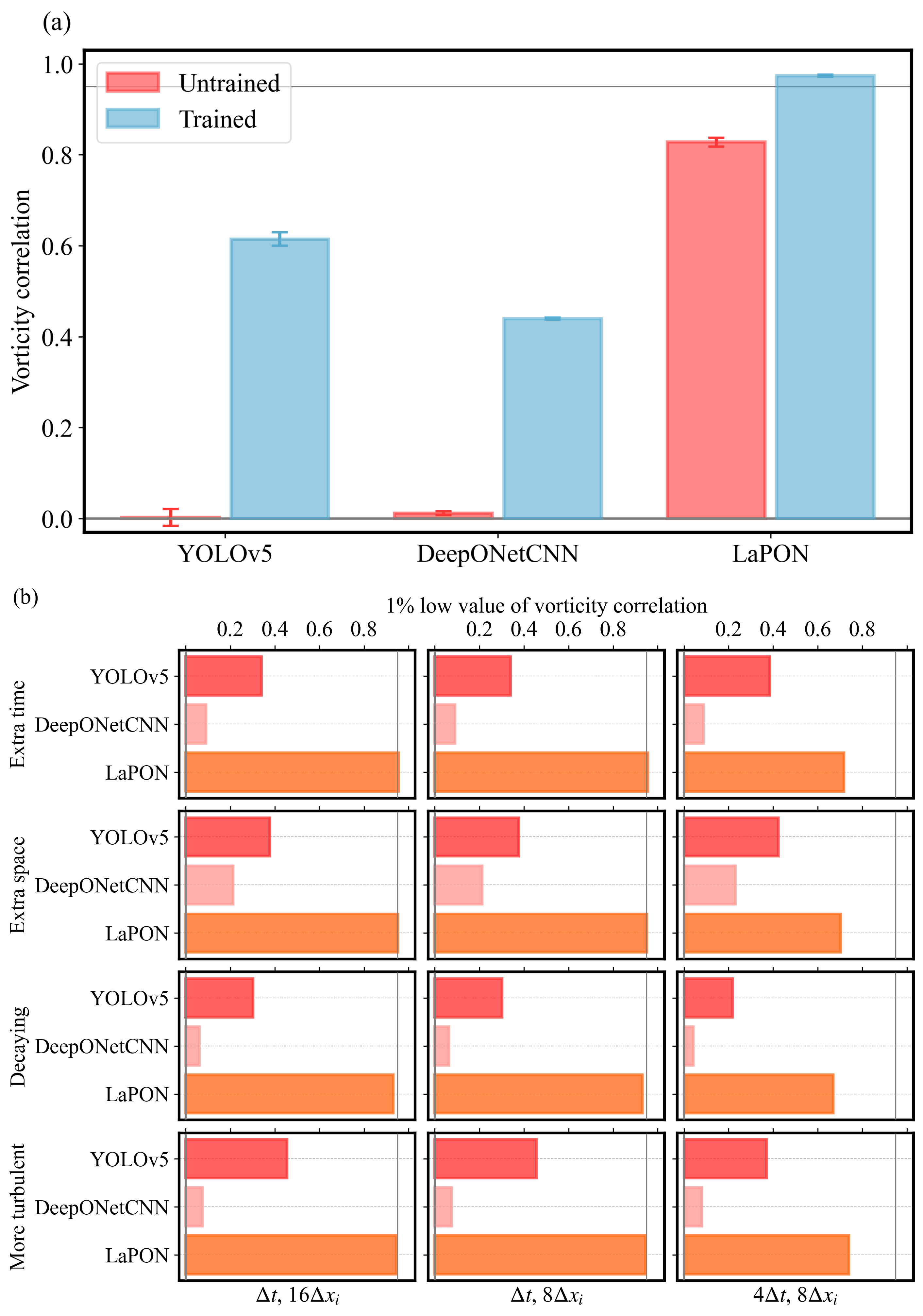}	
	\caption{LaPON outperform two classic ML baseline methods in terms of accuracy, stability and generalization (the evolution time used in the test is about 210 original time steps). (a) The mean and standard deviation of vorticity correlation for each architecture on the training dataset using $8 \times $ the base time step and $64 \times 64$ spatial resolution. (b) Each row within a subplot shows the performance metric (1\% low value of vorticity correlation) of one architecture. The models are trained on isotropic forced turbulence with a certain operating parameter (same as Section \nameref{sec:eval_acc}) and tested for generalization without retraining on four types of out-of-distribution data (extra time, extra space domain, decaying and more turbulent flows with different fluid properties and forcing types). The three columns of subplots correspond to three different spatial and temporal discrete scales ($\Delta t$ is the base time step, $\Delta x_i$ is the grid size of the original dataset). Under the same training data and configuration, the composite metrics of LaPON on the test set are at least about 1 times better than the two popular ML baseline methods.} 
	\label{fig:ml}
\end{figure}

From a utilitarian perspective, numerical simulation methods have the characteristics of accuracy and generalization capabilities, while existing AI algorithms currently have an absolute advantage only in computing speed. Although our method in this work also performs calculations similar to numerical simulation, all calculations are defined in a complete ANN architecture, which makes the method have a computational efficiency that is consistent with existing AI algorithms and far exceeds that of the typical numerical simulation. In addition, our method can solve PDEs under the conditions of the coarse grid and large time step size, which further increases its computational efficiency advantage. In this case, what we are most concerned about in model evaluation is the accuracy and generalization capabilities of the method; at the same time, maintaining the stability of long-term predictions is also an important consideration for the deduction of physical phenomena. Among them, a key aspect of generalization refers to the model’s capability to maintain accuracy in untrained scenarios, such as simulations involving distinct fluid characteristics, varied Reynolds numbers, or alternative forcing mechanisms, despite being trained on a limited set of flow conditions.

In what follows, we compare the accuracy, stability and generalization of our method with direct numerical simulation baselines for simulations of two-dimensional flow. 

Direct numerical simulation loses predictive fidelity when spatial discretization fails to resolve critical small-scale flow features or temporal steps exceed stability thresholds. In contrast, our method significantly reduces this effect. In order to fully verify the ability of the method to learn to find the mean point described by the Lagrange’s mean value theorem, we use 10 times the original time step ($2 \times 10^{-4} s$) used in the simulation of the DNS dataset as the base time step size and evaluate the algorithm with its multiples. Direct learning on the large time scale can effectively reduce the number of time stepping, which helps to alleviate the error accumulation problem under long simulation time, and can also further speed up the simulation. In addition, the coarse grid data in the evaluation are all obtained by coarsening the original high-resolution data. The model is trained only on continuous flow field snapshots that take less than one large-eddy turnover time. And the generalization verification shows the advantage of the model architecture in terms of training data dependence.

\bmhead{Accuracy}\label{sec:eval_acc}

The scalar vorticity field $\omega=\partial_{x} u_{y}-\partial_{y} u_{x}$ provides a convenient way to quantifying fluid motion in turbulence \cite{boffetta2012two}. Accuracy can be quantified by the Pearson correlation of vorticity fields between ground truth solution and predicted state.

Fig. \ref{fig:acc} shows the difference between the single-frame solution of our model (LaPON $64 \times 64$) and the target solution of fully resolved DNS, and compares it with several direct simulation baselines, where the initial condition does not exceed the range of the training data. Strikingly, in Fig. \ref{fig:acc} (b) and (c), LaPON on a coarse grid exceeds accuracy of direct simulation at $8 \times $ higher spatiotemporal resolution, and the accuracy is also significantly higher on large time steps. Due to the chaotic nature of turbulent flows, the larger the time step, the smaller the vorticity correlation after a single time stepping. In addition, the LaPON achieves multi-scale adaptation because time step-dependent conditional operators are designed in the architecture, while existing autoregressive ML models are often only applicable to a single time step. Fig. \ref{fig:acc} (a) shows the prediction of the vorticity field for two different models: the LaPON $64 \times 64$ matches the reference solution more accurately than the direct simulation baseline of the same spatial resolution, it has significantly fewer errors than the baseline. Fig. \ref{fig:acc} (d) compares the energy spectrum $E(\mathbf{k})=\frac{1}{2}|\mathbf{u}(\mathbf{k})|^{2}$ of LaPON and DNS at different spatial resolutions. As the spatial resolution decreases, DNS has difficulty capturing high-frequency features. However, the LaPON that has learned how to eliminate discretization errors accurately captures the energy distribution within the effective spectrum.

\bmhead{Stability}

Although ML models are usually trained for single-step inference, in most cases, we want to get the evolution of the flow field over a period of time, which requires ML models to simulate for a long time like CFD solvers instead of just one step. Therefore, stability is crucial when solving continuously (a kind of regression prediction) at multiple time steps. In order to examine the model's ability to simulate long simulation times, in Fig. \ref{fig:stb}, we further test the stability of the model trained in Section \nameref{sec:eval_acc} on the same data from multiple aspects: the change in accuracy in the time dimension and the standard deviation of accuracy in the sample dimension when solving continuously on multiple time steps; the robustness to noise and incomplete flow field data, where the resistance to noise can help to determine the error accumulation of the model during time stepping.

Compared to DNS, we found that our model can maintain accuracy and stability when time stepping on the coarse grid and the large time step size, even if the Courant–Friedrichs–Lewy condition (the CFL condition) is not satisfied. This is mainly reflected in the following: in the continuous time stepping test in Fig. \ref{fig:stb} (b), although only the training of single-step inference is performed, the performance of LaPON is still higher than the DS baseline of the same spatial resolution; and when using large time step size for continuous time stepping, there is no significant accuracy degradation and standard deviation increase like DS baselines.

In addition, in Fig. \ref{fig:stb} (c) and (d), we found that our model has a higher tolerance to data noise and partial data missing, which may be a general advantage of using ANN models for flow field calculations, because ANN models are usually designed as bottleneck topology structures.

Finally, some other observations: (1) For coarse grids, too small a time step size will increase the number of steps, and the final calculation result of the flow field is more likely to be destroyed by the numerical diffusion. Moreover, although our model is only trained on a larger time step size, its test on a smaller time step size still shows some generalization. (2) For large time step sizes, since the CFL condition is no longer satisfied (which is bad for DNS), the DS baselines with relatively fine grids will have a significant drop in accuracy during time stepping due to the large Courant number. Although our model is also based on explicit time stepping, it can always maintain the stability of the solution. In general, compared with DNS, LaPON on large time step size and coarse grid not only has less computation, but also has considerable improvements in accuracy and stability.

\bmhead{Generalization}

Practical utility of a machine learning model depends on its ability to maintain predictive accuracy for flow regimes beyond the training data domain, even without requiring additional parameter updates. This generalization capacity stems from the model’s focus on extracting fundamental features of the function. We expect our model to generalize well because it learns fundamental operators: what is corrected is not a specific flow field or a specific term in PDE, but the derivative of the function, which is a fundamental feature. As a result, these operators can theoretically be applied to any flow with fundamental partial derivatives similar to those seen during training. 

We consider four different types of generalization tests (see Fig. \ref{fig:main} (a)): (1) extra time (outside the time coordinate range of the training data), (2) extra space domain outside the training distribution (outside the space coordinate range of the training data), (3) unforced decaying turbulent flow with different fluid properties, and (4) Kolmogorov flow with different fluid properties at a larger Reynolds number.

First, we tested generalization on flows under the same forcing but with temporal and spatial domains outside of the training distribution (extrapolation in time and space). Our ML model still achieves high performance because its corrections only act on fundamental features of the flow field function, and the two test flow fields have consistent fluid properties and forcing with the training flow field (See Fig. \ref{fig:ml} (b)). 

In addition, we apply the model to two types of turbulence \cite{chandler2013invariant, kochkov2021machine} with different properties than the training flow. One is a decaying turbulence without forcing that evolves in time from a random initial condition with a high wavenumber component, and the scale of the eddies and the Reynolds number vary with time. The other is a forced turbulence with Kolmogorov forcing and at a higher Reynolds number than the training flow. These two generalization tests are harder: can the model be applied to a turbulence that is significantly nonstationary in time? And can it be applied to a turbulence with a higher Reynolds number where the spatial structure of the eddies is more complex? However, since our model does not learn to fit dynamic characteristics of a specific flow field as most existing end-to-end ML methods do, but rather learns to modify fundamental features of the flow field function, we may not have to face the above difficulties. In Fig. \ref{fig:ml} (b), our model without retraining indeed still maintains high performance. This level of generalization is remarkable because we are now testing the model on two flows that are very different from the training flow. In Fig. \ref{fig:gnr} we further analyze the decaying turbulence. Fig. \ref{fig:gnr} (b) shows that our model still has a significant improvement in accuracy at large time step sizes compared to the direct simulation baseline of the same spatial resolution. Fig. \ref{fig:gnr} (a) visualizes the vorticity, showing that the standard numerical method may be corrupted due to frequency aliasing and numerical dispersion, which is further verified by the energy spectrum shown in Fig. \ref{fig:gnr} (c).

\section*{Discussion}

\subsection*{Comparison to other ML models}

In this section, we compare the performance of LaPON to alternative ML-based methods. We consider two popular and representative ML methods: the classic ML model yolov5 \cite{glenn2022yolov5} and DeepONetCNN \cite{lu2021deeponet, luo2023cfdbench} specifically used to solve PDE. By construction, these models can perform basic explicit time stepping. Each model is trained on the same training dataset described previously, and evaluated on the same generalization task. We compare their performance using the following metrics: the mean and standard deviation of the vorticity correlation between the model prediction and the reference solution to evaluate the accuracy and stability of the models before and after training on the training set; the 1\% low value of vorticity correlation to more conveniently evaluate the accuracy and stability of the models on out-of-distribution (OOD) data.

Fig. \ref{fig:ml} compares the results from several aspects. Overall, we find that LaPON performs the best on different temporal and spatial discrete scales. An impressive and particular strength of our model is its generalization to other flow fields, as seen from its remarkably consistent high performance across different datasets in Fig. \ref{fig:ml} (b). In contrast, end-to-end ML methods do not perform well on OOD data and their performance is not stable across different data. Among them, yolov5 has a relatively high performance, but due to the limited training data, its metrics are still significantly lower than LaPON. The performance of DeepONetCNN is not satisfactory. In addition to the data dependency problem consistent with yolov5, the turbulence with high-dimensional and multi-scale characteristics may not be friendly to its original architecture, although its stability as shown by the standard deviation looks good in Fig. \ref{fig:ml} (a). 

It is worth noting that in Fig. \ref{fig:ml} (a), LaPON still has high accuracy on training data even without any training. This is because numerical calculations are embedded in the architecture of our model, and the pure ANN part to be trained only acts as a basic operator to perform correction functions on the fundamental positions obtained by numerical calculations. In addition, the hard constraints are imposed on the AI correction operators to prevent the correction results from deviating too much from the rough solution given by the low-order numerical calculations. It ensures that the lower limit of accuracy of the model can always be guaranteed. In other words, untrained LaPON is equivalent to a numerical simulation method plus a little restricted noise. Compared with the end-to-end ML model, embedding knowledge directly into the ANN architecture is quite beneficial for both training and inference.

\subsection*{Conclusions}

In this paper, we propose a knowledge-embedded deep learning method for high-precision evolution of solutions to any time-dependent PDEs, which achieves accuracy and generality similar to traditional numerical methods, but with much coarser spatial resolution and time step. The method learns two fundamental operators for partial derivative residuals, and in tests on NSE and turbulence problems, it is close to the accuracy of DNS running at 16 times finer spatial resolution and 160 times finer time step, and the cost of execution is much lower than traditional numerical methods. The method embeds the discrete form of NSE into the ANN architecture, so the method inherently contains the ability to predict the dynamic characteristics of the fluid, even if the model has not been trained. At the same time, we embed the physical prior knowledge of the rotational symmetry into the pure AI module by using a new type of basic convolution operation (the isotropic convolution). In addition, the conditional operators based on the attention mechanism and the hard constraints on the operator outputs make this method naturally capable of adapting to different conditional parameters and maintaining a high lower limit of accuracy. Because of these characteristics, the method significantly generalizes much better than end-to-end black-box ML methods, not only to different spatiotemporal locations (no loss of accuracy when extrapolating) but also to different parameter regimes (Reynolds numbers, etc.) and different forcing functions.

In summary, our method embeds prior knowledge into the model architecture of the neural network without destroying the learnability of the ML model. Compared to classical ML models, they can naturally and rigorously satisfy a given constraint without training. This method significantly improves the accuracy and generalization while greatly improving the computational efficiency of numerical simulations. With the help of AI to accelerate the solution of PDEs, expensive simulations can be solved faster or accuracy can be improved without increasing costs.  In applications ranging from aerodynamic design of aerospace systems to real-time weather prediction, this paradigm shift mitigates traditional bottlenecks imposed by exorbitant time and hardware requirements. Looking ahead, the synergistic integration of domain-specific constraints into ML architectures is poised to drive the development of universal, data-efficient computational frameworks. Such advancements will facilitate robust modeling of multi-physics interaction, potentially transforming fields where computational limitations currently hinder scientific progress.

\section*{Methods}

\subsection*{Nondimensionalization}\label{sec:nondimensionalization}

Corresponding to the dimensionless form of PDEs in the network architecture, we also need a set of matching physical quantity nondimensionalization and inverse nondimensionalization network modules. The former is used to process input data, and the latter acts on output data to restore its distribution. At the same time, nondimensionalization of input data (scaling the data distribution to be close to the standard distribution) has a very important positive effect on the generalization of DL models and the convergence of the optimization process.

The focus of nondimensionalization is the calculation of relevant characteristic physical quantities. Taking NSE as an example, for turbulence problems, the characteristic physical quantities used to scale velocity, length, time, force and pressure, as well as the Reynolds number in this work are as follows:
\begin{gather}
    Total \ kinetic \ energy: \ E_{\text {tot}}=\frac{1}{2}\left\langle u_{i} u_{i}\right\rangle \notag \\
    Rms \ velocity \ (characteristic velocity): \ u^{\prime}=\sqrt{\left(\frac{2}{3}\right) E_{\text {tot}}} \notag \\
    Taylor Micro.Scale (characteristic length):  \lambda=\sqrt{\frac{15 \nu u^{\prime 2}}{\varepsilon}} \notag \\ 
    characteristic \ time: \ T=\lambda / u^{\prime} \notag \\
    characteristic \ force: \ F=\frac{u^{\prime}}{T} \notag \\
    characteristic \ pressure: \ P=\rho u^{\prime 2} \notag \\
    Taylor-scale \ Reynolds: \ Re_{\lambda}=\frac{u^{\prime} \lambda}{\nu}
\end{gather}
where $u$ is the velocity field, $\nu$ is the viscosity, $\varepsilon$ is the dissipation, and the density $\rho$ is a constant.

In addition, we also added a nondimensionalization layer and a corresponding inverse nondimensionalization layer at the input and output of the neural operator. The former is used to improve the convergence and stability of neural network optimization and the robustness during inference. The latter is used to restore the distribution of data.

It is worth noting that when optimizing the model, disabling inverse nondimensionalization may be a better choice (directly training on the calculation results that have not been inverse nondimensionalized), which helps to avoid the gradient of the model's learnable parameters being too low (gradient dispersion, or even overflowing downward). At this point, we only need to do the same nondimensionalization on the sample labels, and the model can still be optimized normally.

\subsection*{Conditional neural operators}

The main research object in the field of computer vision is the 2D image data, which is completely consistent with the 2D physical field data under the regular grid in form. In addition, both usually focus on solving the problem of extracting spatial information. Therefore, it is very beneficial to borrow the advanced model architecture in the field of computer vision as the basic architecture of our neural operator.

Generally speaking, dense physical field data, like image data, has too much redundant information. We need to extract important information from the original data and construct a low-dimensional representation to better process the data. Therefore, the mature and powerful Latent Diffusion Models architecture \cite{rombach2022ldm} is a good choice. The bottleneck AutoEncoder module and the U-Net module in this architecture can efficiently compress information. Processing data in latent space can make the model more stable and reduce the computational cost. In addition, the cross-attention mechanism used to implement conditional image generation tasks can integrate conditional information into the data to be processed, thereby effectively controlling the behavior of the model under different conditions. In this work, we use this mechanism to explicitly incorporate the corresponding conditional parameters mentioned above into the neural operator, thereby achieving flexible control of the state of the model under different conditions.

\subsection*{Extracting history frame information}

The ConvLSTM is a classic network architecture used in the DL field to process spatiotemporal sequence data. It has the ability to extract both spatial and temporal information from data. In the time evolution process of PDEs solution, the multi-frame physical field snapshot with continuous time is a typical spatiotemporal sequence, which contains rich dynamic information of historical physical field evolution and possible forcing information. Effectively utilizing this information can improve the accuracy of ML model in solving equations. To this end, this work adds an additional ConvLSTM module to the neural network architecture, and uses the historical dynamic information extracted by it as a set of additional conditional parameters for the neural operator, so as to maximize the utilization of historical information by the model. In order to make the model meet the spatial rotational symmetry, this module is also implemented based on the isotropic convolution.

\subsection*{Solving the PPE}

In the explicit time marching process of NSE under incompressible conditions, it is usually necessary to solve PPE to calculate the pressure term in NSE. Regarding the flow field dataset that does not directly provide pressure data, we have built an additional set of iterative solution modules based on PyTorch in our intelligent solution framework. This module supports efficient execution of multiple classic iterative methods (such as SOR, Jacobi, Gauss Seidel, etc.) on the GPU. It is worth noting that the calculations in this module are all designed as the in-place calculations, otherwise GPU memory overflow may occur during multiple iterations.

\subsection*{Other details}

\bmhead{A simple proof for spatial operator constraint interval}\label{app:proof}

Here we will give a proof that the constraint interval $[I_L, I_H]$ imposed on the spatial operator contains the exact value of the derivative $f^{\prime}_x(t, x)$ based on the assumptions proposed in the previous article.

When $f^{\prime * (n)}_x(t_0, x_0) \ge f^{\prime * (n+1)}_x(t_0, x_0)$, that is, $r_n \le 0$, let $I_H = f^{\prime * (n)}_x (t_0, x_0)$, $R=| f^{\prime * (n+1)}_x (t, x) - f^{\prime * (n)}_x(t, x) |=| r_n – r_n+1 | \approx | r_n |$, then $I_L = I_H -2R$.

Therefore, $f^\prime_x(t, x)= f^{\prime*(n)}_x(t, x)+ r_n \le f^{\prime*(n)}_x(t, x)= I_H$, and $I_L \approx f^{\prime*(n)}_x(t_0, x_0)-2| r_n |= f^{\prime*(n)}_x(t_0, x_0)+2r_n \le f^{\prime*(n)}_x(t_0, x_0)+r_n =f^\prime_x(t, x)$.

The same applies when $f^{\prime*(n)}_x(t_0, x_0)  \le f^{\prime*(n+1)}_x(t_0, x_0)$.

In summary, $f^\prime_x(t, x) \in [I_L, I_H]$.

\bmhead{Model training}

Our model training is based on the same set of empirical configurations and fine-tuned on them: the optimizer is AdamW, the initial learning rate is set to 1e-4, the loss function is the Smooth L1 Loss, the nominal batch size is 128, and the epochs are not less than 500. Our main data augmentation method is to randomly flip the flow field snapshots horizontally and vertically with a probability of 50\%.

In the training engine, some popular modern tricks are added: cosine learning rate scheduler with 0.01 descent rate, learning rate warmup at the beginning of training, the gradient accumulation, the automatic mixed precision (AMP) training, and the exponential moving average (EMA) model, etc.

\bmhead{Initial and boundary conditions}

Similar to traditional numerical methods, our method also performs time-marching of the flow field based on initial and boundary conditions (I/BCs). Efficiently encoding I/BCs into ANNs is a classic problem. We feed IC directly to the network as input data, and BC is imposed by the padding operation. This is not a very novel method but it is effective enough \cite{ren2022phycrnet}.


\backmatter


\section*{Data availability}

All data used in this study are publicly available and can be accessed at the corresponding URL of the database: https://turbulence.idies.jhu.edu/home or https://github.com/google/jax-cfd.

\section*{Code availability}

Source code is available at GitHub (https://github.com/alanZee/lapon-light-open). The training engine follows common practices in the field of deep learning, and other modules include model implementation, isotropic convolutional generation algorithms, and more.








\end{document}